\documentclass[]{aa}
\usepackage{graphicx,hyperref,txfonts}
\usepackage{caption}
\usepackage{amssymb}
\usepackage{pifont}
\usepackage{subcaption}
\usepackage{array}
\usepackage{color}
\newcommand{\com}[1]{\textnormal{#1}}

\begin{document}

\title{Generalised model-independent characterisation of strong gravitational lenses IV: formalism-intrinsic degeneracies}
\titlerunning{Degeneracies in the gravitational lensing formalism}
\author{Jenny Wagner\inst{1}} 
\institute{Universit\"at Heidelberg, Zentrum f\"ur Astronomie, Astronomisches Rechen-Institut, M\"onchhofstr. 12--14, 69120 Heidelberg, Germany\\
\email{j.wagner@uni-heidelberg.de}
}
\date{Received XX; accepted XX}

\abstract{
Based on the standard gravitational lensing formalism with its effective, projected lensing potential in a given background cosmology, we investigated under which transformations of the source position and of the deflection angle the observable properties of the multiple images remain invariant. \com{These observable properties are} time delay differences, the relative image positions, relative shapes, and magnification ratios.   
As \com{they} only constrain local lens properties, we derive general, local invariance transformations in the areas covered by the multiple images. We show that the known global invariance transformations, for example, the mass-sheet transformation or the source position transformation, are contained in our invariance transformations, when they are restricted to the areas covered by the multiple images and when lens-model-based degeneracies are ignored, like the freedom to add or subtract masses in unconstrained regions without multiple images.
Hence, we have identified the general class of invariance transformations that can occur, in particular in our model-independent local characterisation of strong gravitational lenses.
}
\maketitle

\section{Introduction}

Since the discovery by \cite{bib:Falco} that adding a constant mass sheet to the mass density of a gravitational lens model leaves the observable relative image positions and magnification ratios invariant, a lot of other invariance transformations have been found and investigated, for an overview see, for example, \cite{bib:Gorenstein}, \cite{bib:Liesenborgs1}, \cite{bib:Schneider} and references therein. 
All \com{transformations considered in these works} are treated as global invariance transformations of lens properties that are usually applied to a lens model. 

In the course of this paper series -- \cite{bib:Wagner1}, \cite{bib:Wagner2}, \cite{bib:Wagner3} -- we have developed a lens-model-independent characterisation of local lens properties based on the observables of multiple images. 
We find that we can only constrain ratios of derivatives of the lensing potential, and we interpreted this result as the emergence of a local version of the mass sheet degeneracy of \cite{bib:Falco}. Until now, it has been unclear whether the approach is subject to further, yet undetected degeneracies. This \com{is} the same situation lens-model-based approaches are still in, see for example, \cite{bib:Wagner5}, \cite{bib:Williams}, \cite{bib:Wertz}.

In this fourth paper in the series, we start in Section~\ref{sec:prerequisites} with a summary of the assumptions that the standard gravitational lensing formalism is based on, for details see, for example, \cite{bib:SEF}. Subsequently, in Section~\ref{sec:derivations}, we investigate its general class of formalism-intrinsic invariance transformations in the absence of a lens model. 
These transformations are locally confined to the areas covered by the multiple images, and hence, are the invariance transformations that affect our model-independent lens characterisation. 
In Section~\ref{sec:connections}, we split the known invariance transformations mentioned above into their model-based and model-independent parts. Then, we show, how the model-independent parts can be derived from the general formalism-intrinsic invariance transformations developed in Section~\ref{sec:derivations} and how the model-based degeneracies supplement the model-independent ones. A different kind of invariance transformation, not considered here, was investigated in \cite{bib:Schneider2} and \cite{bib:Wagner3}, namely the degeneracy that arises if a lens model is split into two parts, for example, a main (axisymmetric) lens and a perturber. We concluded in \cite{bib:Wagner3} that this degeneracy is broken with additional (non-lensing) information about the main lens or the perturber. Without it, there is no need for splitting the lens into parts, such that this degeneracy can be avoided by considering a single deflection potential.
At last, we conclude in Section~\ref{sec:conclusions} with a summary of the results as found in Sections~\ref{sec:derivations} and \ref{sec:connections} and determine the most general invariance transformations that affect the equations of our model-independent lens characterisation.


\section{Prerequisites of the standard lensing formalism}
\label{sec:prerequisites}

In order to derive the invariant transformations, we \com{rely} on the standard lensing formalism as detailed in \cite{bib:SEF}. This section briefly summarises the prerequisites and assumptions of the lensing formalism that become important for the derivations performed in Section~\ref{sec:derivations}.

\subsection{Assumptions}
\label{sec:assumptions}

We assume that the light propagation from the source to the observer can be described by geometrical optics. The light coming from the source object is deflected by a quasi-stationary, geometrically thin, deflecting mass distribution that is considered as a small mass perturbation on top of the homogeneous background. 
Given these prerequisites, the gravitational lensing effects are described by means of a gravitational lensing potential, also called Fermat potential. It can be interpreted as the time delay of the light propagation from the source to the observer with respect to an undeflected light propagation. The multiple images appear at positions of stationary time delay in accordance with Fermat's principle.

To describe a gravitational lensing configuration, we \com{define} the following quantities in Cartesian coordinates using the notation of \cite{bib:SEF} (see Figure~\ref{fig:time_delay} for a visualisation):
\begin{itemize}
\item $\boldsymbol{y} = (y_1, y_2)$ as the angular position of a background source located at redshift $z_\mathrm{s}$ along the line of sight,
\item $\boldsymbol{x}_i = (x_{i1}, x_{i2})$ as the angular position of the multiple image $i$ at redshift $z_\mathrm{l}$ along the line of sight,
\item $\phi(\boldsymbol{x},\boldsymbol{y})$ as the lensing potential, describing the mapping of the background source to multiple images,
\item $\psi(\boldsymbol{x})$ as the deflection potential, that is,\@ the part of the lensing potential that represents the deflecting mass distribution, which is integrated along the line of sight and projected onto the plane orthogonal to the light propagation where most of the deflecting mass is concentrated; this lens plane is located at redshift $z_\mathrm{l}$ along the line of sight,
\item $\tau$ as the time delay difference between image $i$ and image $j$ due to the different paths that their light rays take from their mutual source position to the observer. 
\end{itemize}

The standard approach assembles the total deflection potential as a superposition of physically detectable masses (as employed, for instance, in light-traces-mass approaches to reconstruct the deflecting mass distribution, see e.g. \cite{bib:Stapelberg} and \cite{bib:Zitrin}). In our approach, however, we do not superpose individual masses but directly use $\psi(\boldsymbol{x})$ encompassing all contributions of all objects in the lens to the deflection potential at position $\boldsymbol{x}$, irrespective of their (observable) origin. 

\subsection{Limits}
\label{sec:limits}

This effective description by a single, projected and integrated deflection potential reaches its limits when one of the assumptions stated in Section~\ref{sec:assumptions} is not fulfilled anymore. For instance, multiple lens planes are required when there are several gravitational lenses of similar masses aligned along the line of sight or, in the case of a black hole, the deflecting mass distribution cannot be modelled as a Newtonian potential embedded in a Robertson-Walker metric anymore. 

While these cases are comparably rare, there are systematic biases due to observational data acquisition that may also require to extend the standard formalism. For instance, rotations caused by the telescope as constrained by \cite{bib:Bacon1} and described for example,\ in \cite{bib:Bacon} can also cause rotations of the multiple images that are not yet accounted for in the formalism.

In Section~\ref{sec:derivations}, we stay within the framework of the standard single-plane lensing formalism. Subsequently, in Section~\ref{sec:SPT}, we will briefly investigate the connection to the source position transformation, \cite{bib:Schneider}, and its interpretation as an invariance transformation in a double-plane lensing formalism.

\section{Derivations of the invariance transformations}
\label{sec:derivations}

\subsection{Time delay difference as a function of the source position and the deflection angle}
\label{sec:time_delay}

The time delay difference $\tau$ between two multiple images of the same source object
\begin{align}
\tau = D \dfrac{(1+z_\mathrm{l})}{c} \Delta \phi \;, \quad D = \dfrac{D_\mathrm{l}D_\mathrm{s}}{D_\mathrm{ls}}
\label{eq:td}
\end{align}
 is proportional to the difference of the lensing potential between these points $\Delta \phi$. $D$ is the distance ratio involving the angular diameter distances between the lens and the observer, $D_\mathrm{l}$, the source and the observer, $D_\mathrm{s}$, and the lens and the source, $D_\mathrm{ls}$, and $c$ denotes the speed of light\footnote{Equation~\eqref{eq:td} is valid, if $\boldsymbol{x}=\boldsymbol{\xi}/D_\mathrm{l}$, in which $\boldsymbol{\xi}$ is the image position given in units of a length. If $\boldsymbol{\xi}$ is not scaled by the measurable $D_\mathrm{l}$ but an arbitrary scale-length, $\xi_0$, the right-hand side has to be multiplied by $\xi_0^2$.}. 
The difference of the lensing potential between the two images $i$ and $j$ is given by
\begin{align}
 \Delta \phi &= \dfrac12 (\boldsymbol{x}_i - \boldsymbol{y})^2 - \psi (\boldsymbol{x}_i) -  \dfrac12 (\boldsymbol{x}_j - \boldsymbol{y})^2 + \psi (\boldsymbol{x}_j) \label{eq:time_delay} \\
&\equiv  \Delta \mathcal{G}(\boldsymbol{y},\boldsymbol{x}_i, \boldsymbol{x_j}) - \Delta \psi (\boldsymbol{x}_i, \boldsymbol{x_j}) \;,
\label{eq:abbreviations}
\end{align}
in which the first term of Equation~\eqref{eq:abbreviations} denotes the geometric part of the lensing potential and the second term denotes the part due to the deflection potential $\psi$.

Assuming $D$ in Equation~\eqref{eq:td} is known, we investigate all invariance transformations of Equation~\eqref{eq:time_delay}, as they will leave the time delay difference between two multiple images invariant. 
On the other hand, Equation~\eqref{eq:td} is used to determine the Hubble constant, $H_0$ as first proposed in \cite{bib:Refsdal}: if the lensing configuration includes multiple images from a time-varying source with measured time delay difference(s), it is possible to solve Equation~\eqref{eq:td} for $H_0$, if $\Delta \phi$ is reconstructed independently, for example, by the extended host of the time-varying source or additional multiple image systems.
In this case, we have to break the class of invariance transformations
\begin{equation}
D \rightarrow D/\lambda \quad \wedge \quad \Delta \phi \rightarrow \lambda \Delta \phi \quad \lambda \in \mathbb{R} \;,
\label{eq:scaling}
\end{equation}
as detailed in \cite{bib:Sonnenfeld}, \cite{bib:Suyu}, and \cite{bib:Xu}, for instance. Assuming that $\tau$ is subject to a bias in its measurement, $\delta \tau$, the $\delta \tau$ can be brought to the right-hand side of Equation~\eqref{eq:td} and incorporated into the transformation of either $D$ or $\Delta \phi$. Hence, the case of a biased measurement of $\tau$ is equivalent to the degeneracies stated in Equation~\eqref{eq:scaling}. In the following, we will fix $D$, meaning that assume that the cosmological model with its parameter values is given, and consider the invariance transformations of Equation~\eqref{eq:time_delay}.

The positions of all multiple images $i=1,...,n$ and the source are related by the lens equation
\begin{equation}
\boldsymbol{y} = \boldsymbol{x}_i - \boldsymbol{\alpha}(\boldsymbol{x}_i) \quad \forall i  \;,
\label{eq:lens_mapping}
\end{equation}
with the deflection angle $\boldsymbol{\alpha}(\boldsymbol{x})=(\alpha_1(\boldsymbol{x}),\alpha_2(\boldsymbol{x}))$ given by
\begin{equation}
\boldsymbol{\alpha}(\boldsymbol{x}) \equiv \nabla_{\boldsymbol{x}} \psi(\boldsymbol{x}) = \left(\dfrac{\partial \psi}{\partial x_1}, \dfrac{\partial \psi}{\partial x_2} \right)\;.
\label{eq:alpha}
\end{equation}
Equation~\eqref{eq:alpha} requires that $\psi$ is differentiable over the domain spanned by the multiple image positions (see $\mathcal{X}$ in Figure~\ref{fig:local_constraints}), which we denote by $\mathcal{X} \subset \mathbb{R}^2$ in the following. Hence, $\psi$ is continuous in the completion of $\mathcal{X}$, denoted by $\overline{\mathcal{X}}$. 

Using Equation~\eqref{eq:lens_mapping}, $\boldsymbol{y}, \boldsymbol{x}$, and $\psi(\boldsymbol{x})$ in Equation~\eqref{eq:time_delay} are not independent, as already noted by \cite{bib:Gorenstein}. Inserting Equation~\eqref{eq:lens_mapping} into the first term of Equation~\eqref{eq:time_delay}, another formulation, as derived in detail in Appendix~\ref{app:alpha}, reads
\begin{align}
\Delta \phi 
&= \dfrac12 \left(|\boldsymbol{\alpha}(\boldsymbol{x}_i)|^2 - |\boldsymbol{\alpha}(\boldsymbol{x}_j)|^2 \right) - \Delta \psi (\boldsymbol{x}_i, \boldsymbol{x_j}) \\
&= \dfrac12 \left( \boldsymbol{x}_i - \boldsymbol{x}_j \right)^\top \left( \boldsymbol{\alpha}(\boldsymbol{x}_i) + \boldsymbol{\alpha}(\boldsymbol{x}_j)\right) - \Delta \psi (\boldsymbol{x}_i, \boldsymbol{x_j}) \label{eq:time_delay_alpha} \\
&\equiv  \Delta \mathcal{G}(\boldsymbol{\alpha},\boldsymbol{x}_i, \boldsymbol{x_j}) - \Delta \psi (\boldsymbol{x}_i, \boldsymbol{x_j}) \;. \label{eq:abbreviations_alpha}
\end{align}
Compared to Equation~\eqref{eq:abbreviations}, Equation~\eqref{eq:abbreviations_alpha} has the advantage that it does not depend on the unobservable source position anymore. The deflection potential and the deflection angle may be subject to degeneracies, yet, there are additional probes of the gravitational potential, for example, like the measurement of the velocity dispersions in the potential well of the deflecting mass distribution, that can break these degeneracies.

Next, we transform $\boldsymbol{y} \rightarrow \tilde{\boldsymbol{y}}$ and $\boldsymbol{\alpha}(\boldsymbol{x}) \rightarrow \tilde{\boldsymbol{\alpha}}(\boldsymbol{x})$ to obtain the transformed quantities $\Delta \tilde{\phi}$, $\Delta \tilde{\mathcal{G}}$, and $\Delta \tilde{\psi}$ in Equations~\eqref{eq:time_delay} and \eqref{eq:time_delay_alpha}.
If the observed time delay difference is supposed to be left unaltered by the transformations of the variables, then 
\begin{equation}
\Delta \tilde{\phi} =\Delta \tilde{\mathcal{G}} - \Delta \tilde{\psi} = \Delta \mathcal{G} - \Delta \psi = \Delta \phi \;,
\end{equation}
 and, thus, 
\begin{equation}
\delta(\Delta \psi) \equiv \Delta \tilde{\psi}  - \Delta \psi = \Delta \tilde{\mathcal{G}} - \Delta \mathcal{G} \equiv \delta(\Delta \mathcal{G})
\label{eq:equation}
\end{equation}
hold. 

\subsection{Transformations of the source position and the deflection angle}
\label{sec:ti_alpha}

We transform the source position and the deflection angle as 
\begin{equation}
\boldsymbol{y} \rightarrow \tilde{\boldsymbol{y}} \equiv \boldsymbol{y} + \delta \boldsymbol{y} \;, \quad \boldsymbol{\alpha}(\boldsymbol{x}) \rightarrow \tilde{\boldsymbol{\alpha}}(\boldsymbol{x}) \;.
\label{eq:transformations}
\end{equation}
The image positions are free to transform as
\begin{equation}
\boldsymbol{x}_i \rightarrow \boldsymbol{x}_i + \delta \boldsymbol{x}_i \; \forall i \;,
\end{equation}
only the relative image positions, as observables, remain invariant. The transformations of $\boldsymbol{y}$ and the $\boldsymbol{x}_i$ look like linear transformations, yet, this is not necessarily the case. $\delta \boldsymbol{y}$ and $\delta \boldsymbol{x}_i$ can be interpreted as any difference between the original and the transformed $\boldsymbol{y}$ and $\boldsymbol{x}_i$. The $\delta \boldsymbol{x}_i$ can be different for different $\boldsymbol{x}_i$, if we include biases in the measurement into the transformation (as also mentioned in Section~\ref{sec:limits}). For example, an inhomogeneous point-spread function can cause a varying $\delta \boldsymbol{x}_i$. In Section~\ref{sec:connections}, we see that for all known, global degeneracies, $\delta \boldsymbol{x}_i$ can only be a constant for all $\boldsymbol{x}_i$. Inserting these most general transformations into the lens equation, we obtain
\begin{equation}
\boldsymbol{y} + \delta \boldsymbol{y} = \boldsymbol{x}_i + \delta \boldsymbol{x}_i - \boldsymbol{\tilde{\alpha}}(x_i) = \boldsymbol{x}_j + \delta \boldsymbol{x}_j - \boldsymbol{\tilde{\alpha}}(x_j) \;. 
\label{eq:delta_x}
\end{equation}
Hence, we can incorporate the transformation of the image positions in the transformation of the deflection angle, such that the original and the transformed lensing equations for two images $i$ and $j$ read
\begin{align}
\boldsymbol{y} &= \boldsymbol{x}_i - \boldsymbol{\alpha}(x_i) = \boldsymbol{x}_j - \boldsymbol{\alpha}(x_j) \;, \label{eq:observable_trafo}\\
\boldsymbol{y} + \delta \boldsymbol{y} &= \boldsymbol{x}_i - \boldsymbol{\tilde{\alpha}}(x_i) = \boldsymbol{x}_j - \boldsymbol{\tilde{\alpha}}(x_j) \;,
\end{align}
and thus, the transformations of the source position can be expressed as the transformation of the deflection angle as
\begin{align}
 \delta \boldsymbol{y} &= \boldsymbol{\alpha}(x_i) - \boldsymbol{\tilde{\alpha}}(x_i)  = \boldsymbol{\alpha}(x_j) - \boldsymbol{\tilde{\alpha}}(x_j) \;.
\label{eq:spt}
\end{align}
Furthermore, we define $\delta \boldsymbol{\alpha}(\boldsymbol{x}) \equiv \tilde{\boldsymbol{\alpha}}(\boldsymbol{x}) - \boldsymbol{\alpha}(\boldsymbol{x})$. Then, the change in the geometric part of the time delay difference is given by
\begin{align}
\delta(\Delta \mathcal{G}) = \dfrac12 \left( \boldsymbol{x}_i - \boldsymbol{x}_j \right)^\top \left( \delta \boldsymbol{\alpha}(\boldsymbol{x}_i) + \delta \boldsymbol{\alpha}(\boldsymbol{x}_j)\right) \;.
\label{eq:delta_G}
\end{align}
Next, we consider the change in the deflection potential and its derivatives due to a transformation in the source position, fulfilling Equation~\eqref{eq:transformations}. Using Equation~\eqref{eq:alpha}, 
$\delta(\Delta \psi)$ is
\begin{align}
\delta(\Delta \psi) &= - \int \limits_{\boldsymbol{x}_i}^{\boldsymbol{x}_j} \mathrm{d} \boldsymbol{x} \, \delta\boldsymbol{\alpha}(\boldsymbol{x}) = \delta \psi (\boldsymbol{x}_i) - \delta \psi (\boldsymbol{x}_j)\;,
\label{eq:delta_psi}
\end{align}
with $\delta \psi(\boldsymbol{x}) \equiv \tilde{\psi}(\boldsymbol{x}) - \psi(\boldsymbol{x})$ corresponding to the definition of $\delta \boldsymbol{\alpha}(\boldsymbol{x})$.  
Combining the results of Equations~\eqref{eq:delta_G} and \eqref{eq:delta_psi}, we obtain the condition
\begin{align}
\delta \psi (\boldsymbol{x}_i) - \delta \psi (\boldsymbol{x}_j) = \int \limits_{\boldsymbol{x}_j}^{\boldsymbol{x}_i} \mathrm{d} \boldsymbol{x} \, \delta\boldsymbol{\alpha}(\boldsymbol{x}) = \left( \boldsymbol{x}_i - \boldsymbol{x}_j \right)^\top \dfrac{\delta \boldsymbol{\alpha}(\boldsymbol{x}_i) + \delta \boldsymbol{\alpha}(\boldsymbol{x}_j)}{2} \;.
\label{eq:invariance_transform}
\end{align}
under which $\Delta \phi$ remains invariant. 

From the viewpoint of the model-independent lens characterisation, Equation~\eqref{eq:invariance_transform} states that the transformations of the deflection potential and the deflection angle are constrained by information at the positions of the multiple images only, because we do not make any assumptions about the deflection angle in $\mathcal{X}$ for the lack of further observations in this region. 
Since we only required $\psi(\boldsymbol{x})$ to be differentiable in $\mathcal{X}$, $\delta \boldsymbol{\alpha}(\boldsymbol{x})$ can be discontinuous. 
Using the Lebesgue integral, $\delta \boldsymbol{\alpha}(\boldsymbol{x})$ is supposed to be an integrable function over $\mathcal{X}$. Hence, the behaviour of the deflection angle outside the region between the multiple images is irrelevant. 
Furthermore, the integral is invariant if $\delta \boldsymbol{\alpha}(\boldsymbol{x})$ is altered by a null-set. 
As a consequence, we are able to insert deflection angles due to point masses at a countable set of infinitely many positions in the region $\mathcal{X}$, when we set up a lens model.

By construction, relating $\boldsymbol{\alpha}(\boldsymbol{x})$ and $\psi(\boldsymbol{x})$ by Equation~\eqref{eq:alpha}, $\boldsymbol{\alpha}(\boldsymbol{x})$ is a conservative vector field and path-independent, such that $\nabla \times \boldsymbol{\alpha}(\boldsymbol{x}) = 0$. Thus, it is not surprising that the integral on the left-hand side of Equation~\eqref{eq:invariance_transform} does only depend on the values of the deflection potential and the deflection angle at $\boldsymbol{x}_i$ and $\boldsymbol{x}_j$, as given by the right-hand side. Exploiting the linearity of the lens equation, it thus suffices to require $\psi(\boldsymbol{x})$ to be differentiable at the positions of the multiple images.

Inserting Equation~\eqref{eq:spt} into Equation~\eqref{eq:invariance_transform}, we obtain 
\begin{equation}
\delta \psi (\boldsymbol{x}_i) - \delta \psi (\boldsymbol{x}_j) = - \left( \boldsymbol{x}_i - \boldsymbol{x}_j \right)^\top \delta \boldsymbol{y}
\end{equation}
being the most general condition to connect the transformation of the deflection potential with a shift of the source position such that the invariance transformation remains exact ( i.e. $\nabla \times \boldsymbol{\alpha} = 0$).

\subsection{Connection to the deflecting mass density distribution}
\label{sec:ti_kappa}

In the derivation of Equation~\eqref{eq:invariance_transform}, we only required that the deflection angle is given as the gradient of the deflection potential at the positions of the multiple images. Now, we investigate additional constraints on $\psi(\boldsymbol{x})$ and $\alpha(\boldsymbol{x})$ to relate the Laplacian\footnote{$L(\cdot)$ is chosen to better distinguish the Laplacian from the difference operator $\Delta$} L$(\cdot)$ of the deflection potential to the deflecting mass density distribution by the Poisson equation
\begin{equation}
\text{L} \left( \psi(\boldsymbol{x}) \right) = 2 \kappa(\boldsymbol{x}) = 2 \dfrac{\Sigma(\boldsymbol{x})}{\Sigma_\mathrm{cr}}\;, 
\label{eq:poisson_equation}
\end{equation}
in which the convergence $\kappa(\boldsymbol{x})$ is the ratio between the two-dimensional, projected surface mass density $\Sigma(\boldsymbol{x})$ and the critical density 
\begin{equation}
\Sigma_\mathrm{cr} = \dfrac{c^2}{4 \pi G} \dfrac{D_\mathrm{s}}{D_\mathrm{ls} D_\mathrm{l}} \;,
\end{equation}
and as such, assumed to be non-negative, $\kappa(\boldsymbol{x}) \ge 0$. 
For this relation to exist, $\psi(\boldsymbol{x})$ must be at least twice differentiable at the positions of the multiple images, because Equation~\eqref{eq:poisson_equation} is supposed to hold point-wise. Yet, searching for (unique) solutions, we have to find a contiguous region in which Equation~\eqref{eq:poisson_equation} holds, so that we assume it is valid in $\mathcal{X}$. In addition, we have to introduce boundary conditions. Otherwise the solution is determined only up to a harmonic function, that is, a function $f(\boldsymbol{x})$ fulfilling the Laplace equation $L(f(\boldsymbol{x})) = 0$ in $\mathcal{X}$. 

Imposing Dirichlet boundary conditions means setting the value of $\psi(\boldsymbol{x})$ at the boundary of $\mathcal{X}$, $\partial \mathcal{X}$. For Neumann boundary conditions, the value of the outer normal of $\psi(\boldsymbol{x})$ at $\partial \mathcal{X}$ is given. Due to the lens equation, 
\begin{equation}
\nabla_{\boldsymbol{x}} \psi(\boldsymbol{x}) = \boldsymbol{x}_i - \boldsymbol{y} 
\end{equation}
at all observable image positions $\boldsymbol{x}_i$, we cannot employ Neumann boundary conditions at $\partial \mathcal{X}$ due to the unobservable source position, so that we restrict our considerations to Dirichlet boundary conditions. We assume, for instance, that we can obtain values of the gravitational lensing potential from different cosmological probes like the velocity dispersions or X-ray that could provide these boundary conditions.

Employing the Lax-Milgram-theorem (see e.g. \cite{bib:Alt}) for continuous $\kappa(\boldsymbol{x})$ in the (open and bounded) domain $\mathcal{X}$, we can find unique solutions $\psi(\boldsymbol{x})$. These are in the intersection of twice-differentiable functions on $\mathcal{X}$ with continuous functions on $\overline{\mathcal{X}}$, $\psi(\boldsymbol{x}) \in C^2(\mathcal{X}) \cap C^0(\overline{\mathcal{X}})$, under Dirichlet boundary conditions, setting $\psi(\boldsymbol{x}) = g(\boldsymbol{x})$ for all $\boldsymbol{x} \in \partial \mathcal{X}$, $g(\boldsymbol{x})$ being a continuous function on $\partial \mathcal{X}$, $g(\boldsymbol{x}) \in C^0(\partial \mathcal{X})$. In this formulation, we obtain unique solutions that are twice continuously differentiable and as such, \com{they} cannot be degenerate by adding additional point mass deflectors at arbitrary points (as discussed in Section~\ref{sec:ti_alpha}). 
Yet, to obtain such solutions, the formulation requires $\kappa(\boldsymbol{x})$ to be continuous in $\mathcal{X}$, which is an information that is not provided by observations, yet a useful working assumption, as seen in Section~\ref{sec:image_degeneracies}. 

Dropping the assumption of continuity for $\kappa(\boldsymbol{x})$, we resort to the weak solution of Equation~\eqref{eq:poisson_equation}. \com{This means that} $\kappa(\boldsymbol{x})$ is not required to be continuous but only in the function space of square-Lebesgue-integrable functions, $L^2(\mathcal{X})$. 

The unique weak solution is found in the Sobolev function space of compact support with square-integrable
second derivatives $H^{1,2}_0(\mathcal{X})$, if we assume that the lens is of compact support, such that $g(\boldsymbol{x}) \equiv 0$ on $\partial \mathcal{X}$.
Compared to the strong solution considered before, the continuity conditions on $\psi(\boldsymbol{x})$ are relaxed to mere Lebesgue-integrability. Then, we can write the deflection potential and the deflection angle as unique solutions of the weak formulation of Equation~\eqref{eq:poisson_equation} as 
\begin{align}
\psi(\hat{\boldsymbol{x}}) &= \dfrac{1}{\pi} \int_{\mathcal{X}} \mathrm{d}^2 \boldsymbol{x} \, \kappa(\boldsymbol{x}) \ln \left| \hat{\boldsymbol{x}} - \boldsymbol{x} \right| \label{eq:potential} \;, \\
\boldsymbol{\alpha}(\hat{\boldsymbol{x}}) &= \dfrac{1}{\pi} \int_{\mathcal{X}} \mathrm{d}^2 \boldsymbol{x} \, \kappa(\boldsymbol{x}) \dfrac{\hat{\boldsymbol{x}}-\boldsymbol{x}}{\left|\hat{\boldsymbol{x}}-\boldsymbol{x} \right|^2} \;.
\label{eq:angle}
\end{align}
Boundary terms including outer normal derivatives on $\partial \mathcal{X}$ vanish due to $g(\boldsymbol{x}) \equiv 0$. $\kappa(\boldsymbol{x}) \in L^2(\mathcal{X})$ leaves the freedom to add point mass deflectors as in Section~\ref{sec:ti_alpha} again. 

\begin{figure}[t]
\centering
  \includegraphics[width=0.27\textwidth]{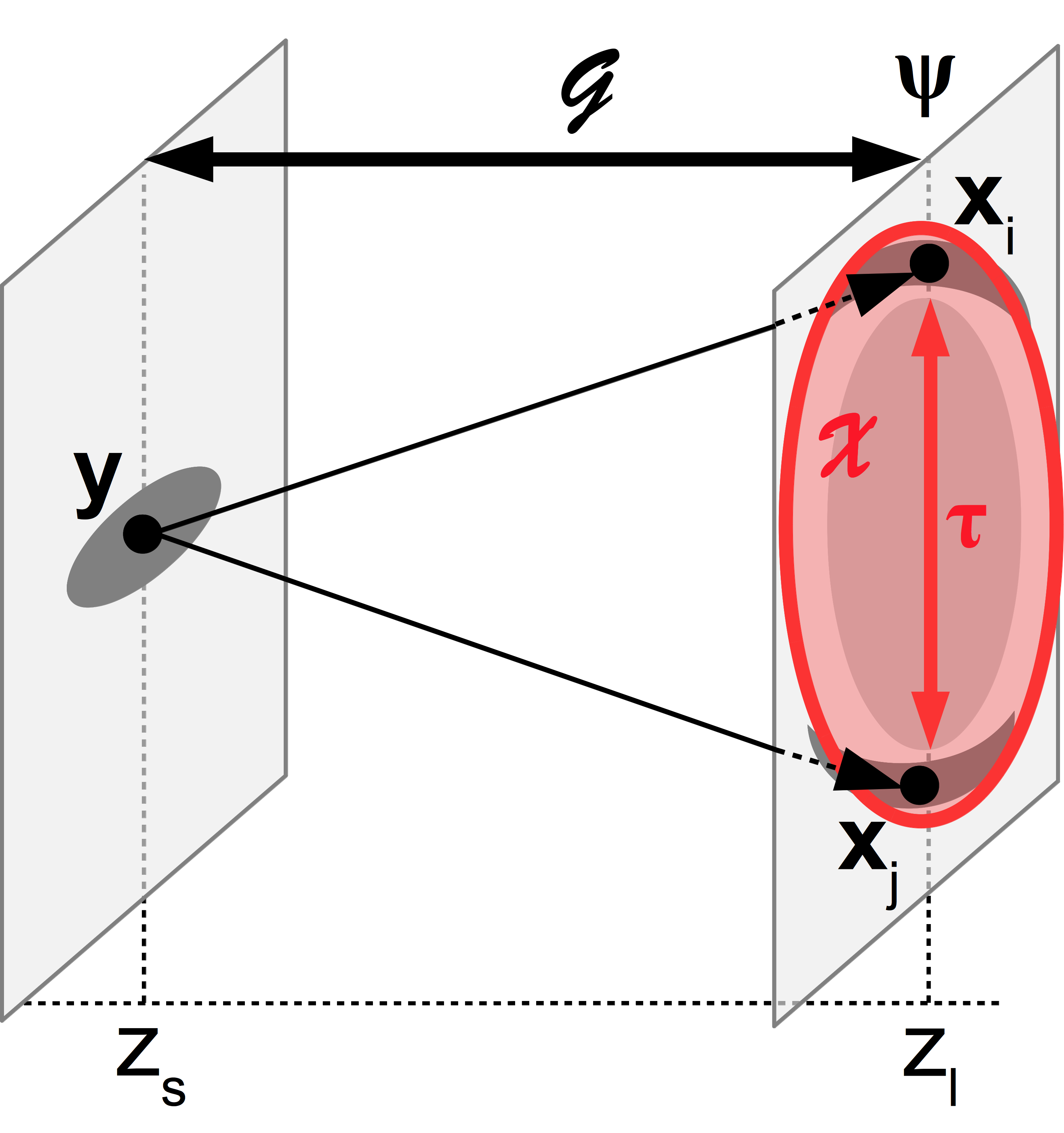}
    \caption{Visualisation of how the observed time delay difference $\tau$ connects the region between the two multiple image positions, $\boldsymbol{x}_i$ and $\boldsymbol{x}_j$ of the source at $\boldsymbol{y}$ in the source plane, as required for Equation~\eqref{eq:poisson_equation} to be uniquely solvable. $\mathcal{G}$ indicates the region of the geometric part of the lensing potential and $\psi$ the plane in which the gravitational lens acts (see Equation~\eqref{eq:abbreviations}).}
\label{fig:time_delay}
\end{figure}

Replacing $\hat{\boldsymbol{x}}$ with the positions of the multiple images and inserting the definitions of Equations~\eqref{eq:potential} and \eqref{eq:angle} for the original and the transformed quantities into Equation~\eqref{eq:invariance_transform}, we arrive at
\begin{align}
\int_\mathcal{X} \mathrm{d}^2 \boldsymbol{x} \, \delta \kappa(\boldsymbol{x}) \, \mathrm{G}_1\left(\boldsymbol{x}_i, \boldsymbol{x}_j, \boldsymbol{x} \right) = \int_\mathcal{X} \mathrm{d}^2 \boldsymbol{x} \, \delta \kappa(\boldsymbol{x}) \, \mathrm{G}_2\left(\boldsymbol{x}_i, \boldsymbol{x}_j, \boldsymbol{x} \right) 
\label{eq:invariance_transform_kappa2}
\end{align}
with $\delta \kappa(\boldsymbol{x})  = \tilde{\kappa}(\boldsymbol{x}) - \kappa(\boldsymbol{x})$ and
\begin{align}
\mathrm{G}_1 \left(\boldsymbol{x}_i, \boldsymbol{x}_j, \boldsymbol{x} \right) &= \ln \left( \dfrac{|\boldsymbol{x}_i - \boldsymbol{x}|}{|\boldsymbol{x}_j - \boldsymbol{x}|} \right) \;, \\
\mathrm{G}_2 \left(\boldsymbol{x}_i, \boldsymbol{x}_j, \boldsymbol{x} \right) &= \dfrac{\left( \boldsymbol{x}_i - \boldsymbol{x}_j \right)^\top}{2}\left( \dfrac{\boldsymbol{x}_i-\boldsymbol{x}}{\left|\boldsymbol{x}_i-\boldsymbol{x} \right|^2} + \dfrac{\boldsymbol{x}_j-\boldsymbol{x}}{\left|\boldsymbol{x}_j-\boldsymbol{x} \right|^2}\right) \;,
\end{align}
as derived in detail in Appendix~\ref{app:ftc}. Then, the two integrands must be equal almost everywhere (i.e. everywhere except for a null-set, as detailed in Section~\ref{sec:ti_alpha}), which implies 
\begin{equation}
\delta \kappa(\boldsymbol{x}) = 0 \quad \Leftrightarrow \quad \tilde{\kappa}(\boldsymbol{x}) = \kappa(\boldsymbol{x}) \quad \text{almost everywhere}\;.
\label{eq:equal_kappa}
\end{equation}

\com{Summarising} our results, we find that a unique solution for $\psi(\boldsymbol{x})$ requires Equation~\eqref{eq:poisson_equation} to hold in a contiguous region connecting the two multiple images with time delay information (see Figure~\ref{fig:time_delay} for a visualisation). It also requires boundary conditions for this region. All local, observable constraints leave $\psi(\boldsymbol{x})$ degenerate up to a harmonic function and allow for arbitrary changes in its values at a null-set of points. The freedom to change the values at a null-set of positions details the idea of introducing point masses into a deflecting mass density stated in \cite{bib:SEF}.
Furthermore, even if we only require a weak unique solution with $\psi(\boldsymbol{x}), \alpha(\boldsymbol{x}) \in H^{1,2}_0(\mathcal{X})$, the deflection angle is bounded and with $\kappa(\boldsymbol{x}) \in L^2(\mathcal{X})$, the deflecting mass is finite, so that we have found weaker constraints on $\psi(\boldsymbol{x}), \alpha(\boldsymbol{x})$, and $\kappa(\boldsymbol{x})$ to fulfil the prerequisites that are necessary to prove the lensing theorems stated in \cite{bib:SEF}. 

Given solutions as Equation~\eqref{eq:potential} and \eqref{eq:angle} exist, Equation~\eqref{eq:equal_kappa} must hold, such that the time delay difference only remains invariant, if the convergence is changed at most at a countable number of positions.
For the strong solution of Equation~\eqref{eq:poisson_equation} with $\psi(\boldsymbol{x}) \in C^2(\mathcal{X}) \cap C^0(\overline{\mathcal{X}})$ and $\kappa(\boldsymbol{x}) \in C^0(\mathcal{X})$, which we will further investigate in the following section, Equation~\eqref{eq:equal_kappa} must hold everywhere. From the observational perspective, these results imply that time delay information tightly constrains the convergence in the region connecting these multiple images.


\subsection{Invariance of further observables under Equation~\eqref{eq:invariance_transform}}
\label{sec:image_degeneracies}

Having first established general invariance transformations of the time delay equation, we now investigate, which other lensing observables remain invariant under these transformations. In the following, we assume the existence of a strong solution with $\psi(\boldsymbol{x}) \in C^2(\mathcal{X}) \cap C^0(\overline{\mathcal{X}})$ and $\kappa(\boldsymbol{x}) \in C^0(\mathcal{X})$ because the standard lensing formalism requires derivatives of $\psi(\boldsymbol{x})$ to exist in the strong (usual) sense and not only in the weak (integral) formulation. Further lensing observables, apart from the relative distances between multiple images, are
\begin{itemize}
\item the magnification ratios between multiple images,
\item the shapes of the multiple images, for example, as a Taylor series expansion into multipole moments, relative to each other,
\item relative distances and shapes of sub-structures, if the resolution of the multiple images is high enough to identify them.\footnote{If sub-structures can be resolved, we consider them as sub-images and apply the following analysis to each sub-image in the same way as to an unresolved multiple image without sub-structure.}
\end{itemize}

The magnification $\mu$ of a multiple image is given as the determinant of the Hessian of $\phi(\boldsymbol{x},\boldsymbol{y})$
\begin{equation}
\mu = \left( \det (A) \right)^{-1} \;,
\label{eq:mu}
\end{equation}
in which $A$ is the magnification (or distortion) matrix
\begin{align}
A = \left( \begin{matrix} 1- \psi_{11} & - \psi_{12} \\ - \psi_{12} & 1-\psi_{22} \end{matrix}\right)
\label{eq:mag_matrix}
\end{align}
and the sub-scripts on $\psi$ denote the partial derivatives in the $x_1$- and $x_2$-direction. From Equation~\eqref{eq:mag_matrix}, we can determine the trace of the Hessian of $\psi(\boldsymbol{x})$ to be $tr(H) = \psi_{11} + \psi_{22}$. Since $\text{L}(\psi(\boldsymbol{x})) = \text{tr}(H)$, a comparison with Equation~\eqref{eq:poisson_equation} yields $\kappa = 1/2 \left( \psi_{11} + \psi_{22} \right)$.

For extended multiply-imaged objects, the shape of the isocontours of their intensity profiles is captured by a multipole expansion of the intensity profile around the centre of light of the image (aligned in a coordinate system with the centre of light of the source). The maximum area that the multipole expansion covers is delimited by neighbouring objects and the signal-to-noise ratio level of the data acquisition. 
Setting the origins of the coordinate systems in the source and lens planes to the centres of light, the Taylor-expanded lens equation 
\begin{align}
y_i = \sum \limits_{j=1}^{2} A_{ij} x_j + \dfrac12 \sum \limits_{j,k=1}^{2} D_{ijk} x_j x_k + \mathcal{O}(\boldsymbol{x}^3)\;, \quad i = 1,2
\label{eq:image_distortions}
\end{align} 
transforms the intensity profile $I(\boldsymbol{y})$ in the source plane to $I(\boldsymbol{x})$ in the lens plane (see e.g. \cite{bib:Goldberg} for details). 
The $A_{ij}$ are the entries of the matrix in Equation~\eqref{eq:mag_matrix} and the $D_{ijk}$ are the third-order partial derivatives
\begin{align}
D_{ijk} = \dfrac{\partial A_{ij}}{\partial x_k}\;, \quad i,j,k = 1,2 \,.
\end{align}
In \cite{bib:Wagner1} and \cite{bib:Wagner2}, we showed that the multipole expansion of the brightness profile of the images approximates the multipole expansion of the lensing potential well and source properties can be neglected, if the multiple images are close to a critical curve and their extension is much smaller than the scale on which $\kappa$ changes.

While the (strong) solution of Equation~\eqref{eq:poisson_equation} only guarantees the existence of second-order derivatives of $\psi(\boldsymbol{x})$, Equation~\eqref{eq:image_distortions} requires $\psi(\boldsymbol{x})$ to be analytic, such that the multipole expansion can be set up and converges to the lens equation (Equation~\eqref{eq:lens_mapping}). 
Irrespective of the question whether higher-order derivatives exist or can be measured, we can state that the so-defined observables are left invariant by the invariance transformations of Equation~\eqref{eq:invariance_transform}. 
This is easy to see: $\delta \kappa (\boldsymbol{x}) = 0$ for $\kappa(\boldsymbol{x}) \in C^0(\mathcal{X})$ implies by Equations~\eqref{eq:delta_x} and \eqref{eq:angle} that $\delta \alpha (\boldsymbol{x})$ is constant in Equation~\eqref{eq:invariance_transform}, so that $\delta \psi(\boldsymbol{x})$ must be linear for all $\boldsymbol{x} \in \mathcal{X}$, such that all derivatives of $\psi(\boldsymbol{x})$ of second-order or higher remain invariant.

\subsection{Invariance transformations without time delay information}
\label{sec:local_gmst}

\begin{figure}[t]
\centering
  \includegraphics[width=0.24\textwidth]{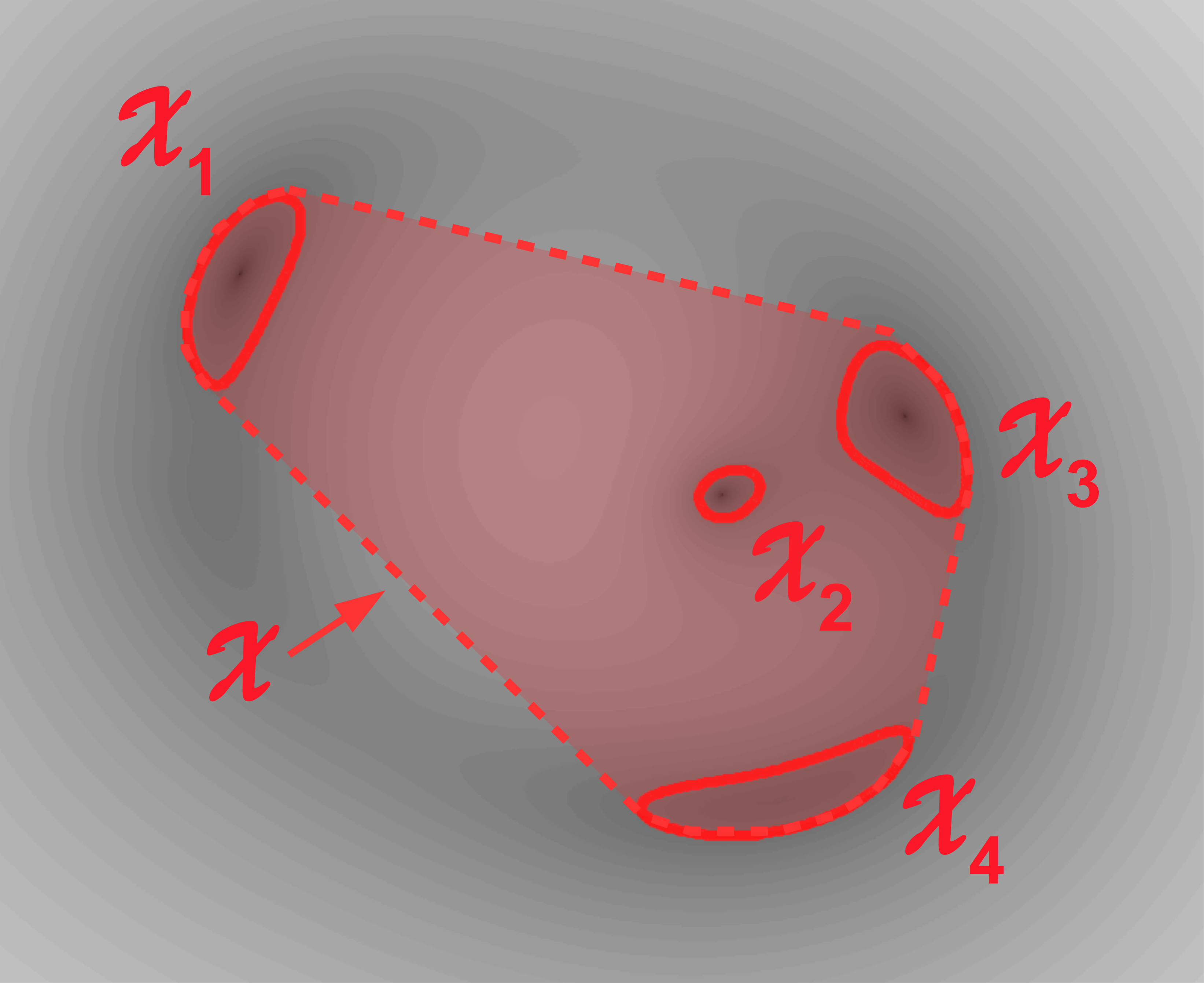}
    \caption{Visualisation of the local regions $\mathcal{X}_i$, $i=1,...,n$ in which Equation~\eqref{eq:poisson_equation} holds in the absence of time delay information. The regions are delineated by the isophotes of the multiple images and the region $\mathcal{X}$ connects all $n=4$ multiple images.}
\label{fig:local_constraints}
\end{figure}

For most multiple image systems, no time delay differences can be measured, so that we only observe the relative positions, shapes, and magnification ratios of the multiple images\footnote{We assume that the multiple images are extended because practically all examples include multiple images of an extended object, for example, a host galaxy or a core and a jet of a quasar observed in radio bands.}. 
Since no time delay information connecting the multiple images with each other is available, we do not require Equation~\eqref{eq:poisson_equation} to hold in $\mathcal{X}$. 
Instead, we only assume that Equation~\eqref{eq:poisson_equation} holds in the (not necessarily connected) areas that are covered by the multiple images around their centres of light, which we will denote by $\mathcal{X}_i$ for each multiple image $i=1, ..., n$. 
If multiple images overlap, we assume their $\mathcal{X}_i$ to be united, otherwise, the $\mathcal{X}_i$ are disjunct with $\overline{\mathcal{X}}_i \cap \overline{\mathcal{X}}_j = \left\{ \right\}$ for all $i \ne j$ (see Figure~\ref{fig:local_constraints} for a visualisation).

Hence, assuming that a solution for $\psi(\boldsymbol{x})$ and $\kappa(\boldsymbol{x})$ exists in each $\mathcal{X}_i$, denoted by $\psi_i(\boldsymbol{x})$ and $\kappa_i(\boldsymbol{x})$, respectively, we insert Equation~\eqref{eq:angle} into Equation~\eqref{eq:spt} to obtain
\begin{align}
\delta \boldsymbol{y} = \delta \boldsymbol{\alpha}(\boldsymbol{x}_i) = \dfrac{1}{\pi} \int_{\mathcal{X}_i} \mathrm{d}^2 \boldsymbol{x}  \, \delta\kappa_i(\boldsymbol{x}) \dfrac{\boldsymbol{x}_i-\boldsymbol{x}}{\left|\boldsymbol{x}_i-\boldsymbol{x} \right|^2} \;, \quad \text{in} \; \mathcal{X}_i \quad \forall i\;,
\label{eq:gmst}
\end{align}
as the general invariance transformation that has to be fulfilled locally in all $\mathcal{X}_i$ in the absence of time delay information, keeping in mind that $\delta \boldsymbol{\alpha}(\boldsymbol{x}_i) = \tilde{\boldsymbol{\alpha}}(\boldsymbol{x}_i) - \boldsymbol{\alpha}(\boldsymbol{x}_i) - \delta \boldsymbol{x}_i$ from Equation~\eqref{eq:delta_x} before absorbing $\delta \boldsymbol{x}_i$ into $\tilde{\boldsymbol{\alpha}}(\boldsymbol{x}_i)$. Here, the boundary terms vanish because we assume that the Dirichlet boundary conditions on $\partial \mathcal{X}_i$ are given by further non-lensing observables, such that they also remain invariant (see Appendix~\ref{app:ftc} for details).
As before, depending on whether the solution is strong or weak, degeneracies at a null-set of positions are evaded or not. 

The change in the distortion matrix consequently reads
\begin{align}
\tilde{A} - A = \left( \dfrac{\partial (\delta \boldsymbol{y})}{\partial \boldsymbol{x}}\right) = \left( \dfrac{\partial (\delta \boldsymbol{\alpha}(\boldsymbol{x}_i))}{\partial \boldsymbol{x}}\right) \;, \quad \text{in} \; \mathcal{X}_i \quad \forall i\;.
\end{align}
If we assume that $\partial (\delta \boldsymbol{y})/\partial \boldsymbol{x} \ne 0$, (the zero matrix), the transformation derived in Equation~\eqref{eq:gmst} only leaves the relative shapes of the multiple images invariant. Furthermore, we obtain
\begin{align}
\text{tr} \left( \tilde{A} - A\right) = -2 \delta \kappa(\boldsymbol{x}_i) \;, \quad \forall i \;,
\label{eq:gmst2}
\end{align}
which gives a point-wise invariance transformation at the positions of the multiple images, analogous to Equation~\eqref{eq:invariance_transform}, only considering its left- and right-hand sides. It is equivalent to Equation~\eqref{eq:poisson_equation} but the right hand side can also be interpreted as the transformation of the brightness moments as linked to the distortion matrix in Section~\ref{sec:image_degeneracies}.

While the time delay information connected the multiple image positions and caused $\kappa(\boldsymbol{x})$ to be strictly constrained, the second parts of Equations~\eqref{eq:gmst} and \eqref{eq:gmst2} leave a lot more freedom: the transformation of $\kappa(\boldsymbol{x})$ is split into disconnected regions $\mathcal{X}_i$, where local $\delta \kappa_i(\boldsymbol{x})$ can only be constrained under the integrals over the individual $\mathcal{X}_i$ or by deriving the transformation difference of the source or the deflection angle with respect to $\boldsymbol{x}$. Example transformations are discussed subsequently in Section~\ref{sec:connections} along with the known global degeneracies.

\section{Connections to known degeneracies}
\label{sec:connections}

In Section~\ref{sec:derivations}, we have determined general, local invariance transformations under which observable properties of multiple images remain invariant. Now, we revise the existing transformations from the literature, connect them with each other, and integrate them into our framework. 
Since they are usually employed in the context of a lens model, they can contain invariance transformations that only affect regions without multiple images, such as the monopole degeneracy detailed in Section~\ref{sec:monopole}. As lens models yield an encompassing description of a lensing configuration in the entire lensing region or even the lens plane, they also aim at establishing a global invariance transformation. Hence, basing on the local invariance transformations of Section~\ref{sec:derivations}, we now show that these global invariance transformations fulfil Equations~\eqref{eq:invariance_transform} or \eqref{eq:gmst} and how the additional (model-based) parts of these invariance transformations can be attached to the local ones. If not mentioned otherwise, we will assume the strong solution of Equation~\eqref{eq:poisson_equation} to exist. 
In addition, we cut the Taylor-series in Equation~\eqref{eq:image_distortions} after the quadrupole moment, since, so far, only these moments have been included in analyses of multiple images, see for example, \cite{bib:Jullo, bib:Liesenborgs2, bib:Wagner1, bib:Wagner2}.

\subsection{Monopole degeneracy}
\label{sec:monopole}

The monopole degeneracy was introduced in \cite{bib:Saha} and is best explained in \cite{bib:Liesenborgs1} as a degeneracy of lens models stating that it is possible to add an axisymmetric mass density of finite extent in areas that do not contain multiple images, such that the observable image positions, shapes and magnification ratios remain invariant. 
The potential outside the region of the additional mass is changed by a constant. As the time delay difference only measures potential differences, it also remains unaffected by the additional mass.

This degeneracy naturally arises in our framework, as established in Section~\ref{sec:derivations}: Equations~\eqref{eq:invariance_transform} and \eqref{eq:gmst} only hold in $\mathcal{X}$ and in the areas covered by the multiple images $\mathcal{X}_i$, $i=1, ..., n$, respectively. Thus, we can add arbitrary mass densities without altering any observable, if the added mass density $\kappa_\mathrm{a}(\boldsymbol{x})$ fulfils the \com{following} constraints:
\begin{itemize}
\item $\kappa_\mathrm{a}(\boldsymbol{x})$ is of finite extent, meaning that it is confined to a region $\mathcal{X}_\mathrm{a}$, such that $\mathcal{X}_\mathrm{a} \cap \mathcal{X} = \left\{ \right\}$, if time delay information is available, and $\mathcal{X}_\mathrm{a} \cap \mathcal{X}_i = \left\{ \right\}$ for all regions $i$ around the multiple image positions, if no time delay information is given. 
\item In order to set up a global lens model in the lens model domain $\mathcal{X}_\mathrm{c}$ containing all multiple images and $\mathcal{X}_\mathrm{a}$ with $\psi_\mathrm{c}(\boldsymbol{x}) \in C^2(\mathcal{X}_\mathrm{c}) \cap C^0(\overline{\mathcal{X}_\mathrm{c}})$ and $\kappa_\mathrm{c}(\boldsymbol{x}) \in C^0(\mathcal{X}_\mathrm{c})$, we have to define the additional mass density with $\psi_\mathrm{a}(\boldsymbol{x}) \in C^2(\mathcal{X}_\mathrm{a}) \cap C^0(\overline{\mathcal{X}_\mathrm{a}})$ and $\kappa_\mathrm{a}(\boldsymbol{x}) \in C^0(\mathcal{X}_\mathrm{a})$, such that the boundary conditions on $\partial \mathcal{X}_\mathrm{a} \cap \partial \mathcal{X}$ (or $\partial \mathcal{X}_\mathrm{a} \cap \partial \mathcal{X}_i$) are chosen accordingly, in case $\partial \mathcal{X}_\mathrm{a} \cap \partial \mathcal{X} \ne \left\{ \right\}$ (or $\partial \mathcal{X}_\mathrm{a} \cap \partial \mathcal{X}_i \ne \left\{ \right\}$). 
\end{itemize}

Hence, we have explained how the monopole degeneracy arises from the fact that multiple images only put local constraints on the deflecting mass \com{or deflection} potential, such that it is always possible to add a mass of finite extent in regions without multiple images without any effect on the observables. In addition, we have generalised the monopole degeneracy because the requirements on $\kappa_\mathrm{a}(\boldsymbol{x})$ to leave the observables invariant do not necessarily include axisymmetry. 
Being only a degeneracy of regions without multiple images in global lens models, the model-independent, local lens characterisation as developed in \cite{bib:Wagner1}, \cite{bib:Wagner2}, and \cite{bib:Wagner3} is not affected by this degeneracy. 

\subsection{Invariance transformations of \cite{bib:Gorenstein}}
\label{sec:Gorenstein}

Starting with the discovery of the mass sheet degeneracy in \cite{bib:Falco}, classes of invariance transformations were found and summarised in \cite{bib:Gorenstein}. 
The transformations are applied to the entire lens plane, meaning that they are global, but, as such are also applied at the multiple image positions. This implies that the generalised, model-independent lens characterisation is also subject to those degeneracies. Their restrictions to the region covered by the multiple images (i.e. to $\mathcal{X}$ or $\mathcal{X}_i$ for all $i$) can be considered as intrinsic invariance transformations of the gravitational lensing formalism. 
In addition, \cite{bib:Gorenstein} investigate the resulting degeneracies of the parameters of global lens models that arise due to these invariance transformations. We briefly summarise the invariance transformations of \cite{bib:Gorenstein} in Table~\ref{tab:Gorenstein}. 

\begin{table}[t]
\caption{Invariance transformations of \cite{bib:Gorenstein}: first two columns: name and transformation law, third column: change in the time delay difference due to the transformation, fourth column: change in the relative image positions, fifth column: change in the magnification ratios; $c, s, \epsilon \in \mathbb{R}$.}
\begin{tabular}{ccccc}
\hline 
Name & Transformation & $\tilde{\tau}$ & $\Delta \tilde{\boldsymbol{x}}$ & $\tilde{\mu}_i / \tilde{\mu}_j$ \\
\hline
prismatic & $\tilde{\boldsymbol{\alpha}} = \boldsymbol{\alpha} + c$ & inv. & inv. & inv. \\
 & $\tilde{\boldsymbol{y}} = \boldsymbol{y} + c$ &  & \\
\noalign{\smallskip}
similarity & $\tilde{D}_\mathrm{i} = s D_\mathrm{i}$ & $s \tau$ & inv. & inv. \\
\noalign{\smallskip}
magnification & $\tilde{\boldsymbol{\alpha}} = \epsilon \boldsymbol{\alpha} - (1-\epsilon) \boldsymbol{x}$ & $\epsilon \tau$ & inv. & inv. \\
 & $\tilde{\boldsymbol{y}} = \epsilon \boldsymbol{y}$ &  & \\
\hline
\end{tabular}
\label{tab:Gorenstein}
\end{table}
\cite{bib:Gorenstein} conclude that a combination of a similarity and a magnification transformation with $s = 1/\epsilon$ can be set up that also leaves the time delay difference invariant. This transformation is the same as the one mentioned in Equation~\eqref{eq:scaling}. 

Except for the similarity transformation that we do not consider, it is easy to incorporate the prismatic and the magnification transformation into our framework as set up in Section~\ref{sec:derivations}. They are special cases of the general invariance transformation defined by Equation~\eqref{eq:spt}, which can be easily proven by inserting the transformations stated in Table~\ref{tab:Gorenstein} into Equation~\eqref{eq:spt}.
Deriving the deflection angle of the magnification transformation, we arrive at the mass sheet transformation, which is thus naturally incorporated in this class of invariance transformations.

\subsection{Generalised mass sheet transformation}
\label{sec:gmst}

In \cite{bib:Liesenborgs1}, the mass sheet degeneracy is generalised, showing that it cannot be broken it with multiple images from more than one source at different redshifts. 
Instead of adding a global constant in the entire lens plane, \cite{bib:Liesenborgs1} note that it is possible to define a global, generalised mass sheet degeneracy that first scales $\kappa(\boldsymbol{x})$ by all mass sheet transformations of the individual multiple image systems. 
Subsequently, they add monopoles, as introduced in Section~\ref{sec:monopole}, to compensate the effects of the mass sheets for each multiple image system from one source at the positions of the multiple images of all other sources. As a result, each multiple image system is only subject to its own mass sheet transformation.

The derivations in Section~\ref{sec:derivations} describe local properties of the lens around the multiple images, so that we can restrict the mass sheet transformations for each system of multiple images of \cite{bib:Liesenborgs1} to the vicinity of the multiple images analogously to the partition of the lens plane we made in Section~\ref{sec:local_gmst}. Then, we see that, without time delay information, we can construct a global, non-constant invariance transformation that reduces to the individual mass sheet transformations of the multiple image systems in the vicinity of the multiple images. Hence, as already found in \cite{bib:Wagner1} and \cite{bib:Wagner2}, the model-independent local lens characterisation is subject to this local mass sheet degeneracy. As no global connecting lens model is assumed in this approach, no additional monopoles are involved in the local version of the mass sheet transformation.


\subsection{Source position transformation}
\label{sec:SPT}

In \cite{bib:Schneider}, \cite{bib:Unruh}, and \cite{bib:Wertz}, the so-called source position transformation (SPT) is developed and its effects on the observable multiple image properties are investigated.  
The SPT also starts with Equation~\eqref{eq:observable_trafo} for all images of the same source. 
Any transformation must obey the same equation for the transformed deflection angle, $\tilde{\boldsymbol{\alpha}}(\boldsymbol{x})$ and the transformed source position, $\tilde{\boldsymbol{y}}$
\begin{align}
\boldsymbol{y} = \boldsymbol{x}_i - \boldsymbol{\alpha}(\boldsymbol{x}_i) =  \boldsymbol{x}_j - \boldsymbol{\alpha}(\boldsymbol{x}_j) \;, \\
\tilde{\boldsymbol{y}} = \boldsymbol{x}_i - \tilde{\boldsymbol{\alpha}}(\boldsymbol{x}_i) =  \boldsymbol{x}_j - \tilde{\boldsymbol{\alpha}}(\boldsymbol{x}_j)  \;,
\end{align}
such that the original and the transformed equation are linked by
\begin{align}
\tilde{\boldsymbol{\alpha}}(\boldsymbol{x}) = \boldsymbol{\alpha}(\boldsymbol{x}) + \boldsymbol{y} - \tilde{\boldsymbol{y}} = \boldsymbol{x} - \tilde{\boldsymbol{y}}(\boldsymbol{x}-\boldsymbol{\alpha}(\boldsymbol{x})) \;.
\label{eq:spt_alpha}
\end{align}
From this, \cite{bib:Schneider} state that any SPT transforms $\tilde{\boldsymbol{y}}(\boldsymbol{y})$ as a global transformation of the (unobservable) source plane. It gives rise to a deflection angle $\tilde{\boldsymbol{\alpha}}(\boldsymbol{x})$ that keeps existing multiple image positions invariant and does not cause new multiple images to appear. This implies that $\det \left(\partial \tilde{\boldsymbol{y}}/\partial \boldsymbol{y}\right) \ne 0$ must hold in the region of the source plane that covers at least the source positions of the convex hull of multiple images to which the SPT is applied. As a consequence, the SPT-transformed magnification matrix $\tilde{A}$ reads
\begin{align}
\tilde{A} = \left( \dfrac{\partial \tilde{\boldsymbol{y}}}{\partial \boldsymbol{x}} \right) = \left( \dfrac{\partial \tilde{\boldsymbol{y}}}{\partial \boldsymbol{y}} \right)  \left( \dfrac{\partial \boldsymbol{y}}{\partial \boldsymbol{x}} \right) = \left( \dfrac{\partial \tilde{\boldsymbol{y}}}{\partial \boldsymbol{y}} \right)  A \equiv B \cdot A \;,
\label{eq:spt_A}
\end{align}
such that the relative magnification ratios and relative shapes of the multiple images remain invariant (see \cite{bib:Schneider} for details). \cite{bib:Schneider} find that, in general $\tilde{A}$ need not be a symmetric matrix, that is, the SPT-transformed deflection angle $\tilde{\boldsymbol{\alpha}}(\boldsymbol{x})$ need not be the derivative of a deflection potential $\psi(\boldsymbol{x})$, as we required in Section~\ref{sec:time_delay}. As a consequence, $\nabla \times \tilde{\boldsymbol{\alpha}}(\boldsymbol{x}) \ne 0$ is possible and the works of \cite{bib:Schneider}, \cite{bib:Unruh}, and \cite{bib:Wertz} investigate such cases in detail. SPTs obeying $\nabla \times \tilde{\boldsymbol{\alpha}}(\boldsymbol{x}) = 0$ are either the standard mass sheet transformation or transformations of axisymmetric lens models.  

\subsubsection{SPTs with $\nabla \times \tilde{\boldsymbol{\alpha}}(\boldsymbol{x}) = 0$}
Since we only considered deflection angles with $\nabla \times \tilde{\boldsymbol{\alpha}}(\boldsymbol{x}) = 0$ in Section~\ref{sec:derivations}, we first integrate this class of SPTs into our framework: 
the exact SPT transformation of an axisymmetric lens, as stated in \cite{bib:Schneider},
\begin{align}
\tilde{r_y} = (1 + f(r_y)) r_y
\label{eq:SPT_t1}
\end{align}
with an even function $f(r_y)$ expressed in polar coordinates for $\boldsymbol{y} = (r_y, \varphi_y)$ and $\boldsymbol{x} = (r, \varphi)$ gives rise to the transformed (radial) deflection angle
\begin{align}
\tilde{\alpha}(r) = \alpha(r) - (1 + f(r_y)) r_y \;,
\label{eq:SPT_t2}
\end{align}
for a one-to-one SPT.
Inserting $\delta r_y = f(r_y) r_y$ from Equation~\eqref{eq:SPT_t1} and Equation~\eqref{eq:SPT_t2} into Equation~\eqref{eq:spt}, we see that this special SPT fulfils this invariance transformation of our framework.

As calculated in \cite{bib:Wertz}, the difference between the SPT-transformed and the original time delay difference is given by
\begin{align}
\tilde{\tau} - \tau &= - f(r_y) r_y (r_i - r_j) - \delta \psi(r_i) + \delta \psi(r_j)  \\
 &= - f(r_y) r_y (r_i - r_j) + \int \limits_{r_j}^{r_i} f(r_y) r_y \mathrm{d} \boldsymbol{x} \;.
\end{align}
We obtain the same equation when we subtract the time delay differences $\tilde{\tau} = D(1+z_\mathrm{l})/c \Delta \tilde{\phi}$ and $\tau = D(1+z_\mathrm{l})/c \Delta \phi$, using the lensing potentials and their components as defined in Equations~\eqref{eq:delta_G} and \eqref{eq:delta_psi} with $\delta \alpha (r) = -(1+f(r_y)) r_y$. But instead of approximating the change in the time delay difference with respect to the SPT, as done in \cite{bib:Wertz}, we treat the time delay difference as an observable and employ Equation~\eqref{eq:invariance_transform} to state that 
\begin{align}
(r_i - r_j) f(r_y) r_y &= \int \limits_{r_j}^{r_i} f(r_y(r)) r_y(r) \mathrm{d} r \\
\Leftrightarrow (r_i - r_j) \dfrac{(\delta \alpha(r_i)+\delta \alpha(r_j))}{2} &= \int \limits_{r_j}^{r_i} \delta \alpha(r) \mathrm{d}r
\end{align}
leaves the time delay difference invariant.

\subsubsection{SPTs with $\nabla \times \tilde{\boldsymbol{\alpha}}(\boldsymbol{x}) \ne 0$}

\begin{figure}[t]
\centering
  \includegraphics[width=0.38\textwidth]{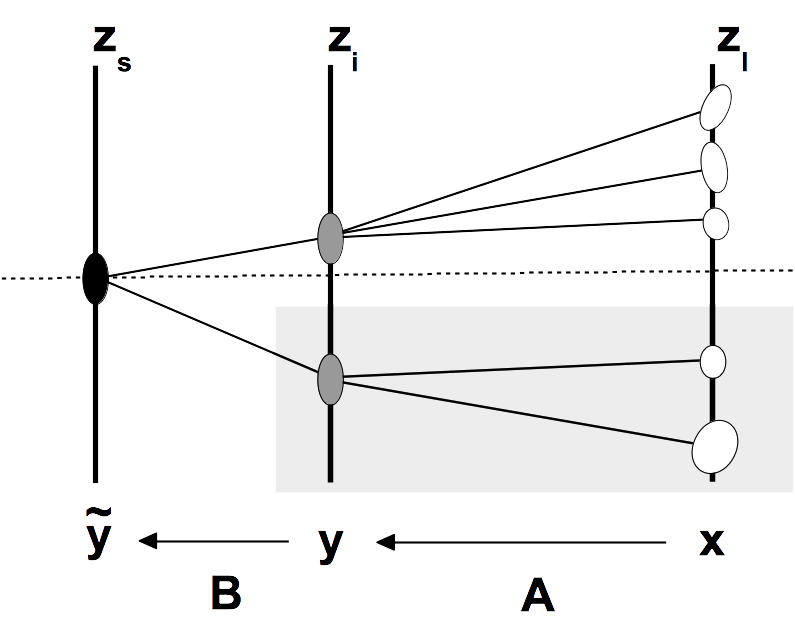}
    \caption{Instead of interpreting $\tilde{\boldsymbol{y}}(\boldsymbol{y})$ as a global transformation in the source plane, it can be viewed as a one-to-one lens mapping, given that the (theoretically possible) multiple image system in the grey-shaded box is not observed.}
\label{fig:spt}
\end{figure}

\noindent
In principle, $\nabla \times \tilde{\boldsymbol{\alpha}}(\boldsymbol{x}) \ne 0$ can arise for several reasons. As stated in \cite{bib:SEF}, a moving, deflecting mass distribution can be a possible cause. Yet, this effect scales with $v/c$, $v$ being the velocity of the moving mass distribution, such that it is, for instance, three orders of magnitude smaller than the lensing effect coming from the deflection potential $\psi(\boldsymbol{x})$ of a galaxy. Secondly, \cite{bib:Bacon} consider systematic, not necessarily gravitational effects to second and third order, for example, a rotation of the telescope that leads to an asymmetric distortion matrix like Equation~\eqref{eq:spt_A}. Third, we can interpret Equation~\eqref{eq:spt_A} as a multi-lens-plane lensing effect as drawn in Figure~\ref{fig:spt}. Instead of a global transformation of the source plane, we can treat $\tilde{\boldsymbol{y}}(\boldsymbol{y})$ as a first, local lens mapping of a source at $\tilde{\boldsymbol{y}}$ to $\boldsymbol{y}$ in a lens plane at an intermediate redshift $z_\mathrm{i}$ with $z_\mathrm{l} < z_\mathrm{i} < z_\mathrm{s}$, which is followed by a second, local lens mapping $\boldsymbol{y}(\boldsymbol{x})$ to the lens plane at $z_\mathrm{l}$. If we require the sequence not to generate any new multiple images, the first lens mapping has to be one-to-one with $\det(B) \ne 0$ for all $\boldsymbol{y}$, such that potentially additional multiple images indicated in the grey box in Figure~\ref{fig:spt} cannot occur. 

In this picture, $\nabla \times \tilde{\boldsymbol{\alpha}}(\boldsymbol{x}) \ne 0$ and an asymmetric $\tilde{A}$ arise if $A$ and $B$ cannot be diagonalised simultaneously, which physically means that there is no effective single-lens-plane description for the double-plane lensing. Since, we consider only single-plane lensing, this case does not occur in the model-independent lens characterisation by construction and is not treated in our framework described in Section~\ref{sec:derivations}. However, interpreting the SPT as lensing by two lens planes, the transformation $\tilde{\boldsymbol{y}}(\boldsymbol{y})$ is not required to be a global transformation anymore. Then, the SPT becomes equivalent to Equation~\eqref{eq:gmst} if $\nabla \times \tilde{\boldsymbol{\alpha}}(\boldsymbol{x}) = 0$.

Summarising the results, we find that the generalised version of the magnification transformation as expressed by Equation~\eqref{eq:spt_alpha} is equivalent to our Equation~\eqref{eq:spt}. 
If we ask for the invariance transformations of SPTs with $\nabla \times \tilde{\boldsymbol{\alpha}}(\boldsymbol{x}) = 0$ for the time delay difference, we arrive at the same results as in Equation~\eqref{eq:invariance_transform}. 
Yet, in our framework, Equation~\eqref{eq:invariance_transform} is not limited to cases with axisymmetry or global transformations because our framework treats local constraints provided by the observables which are, in addition, assumed to be invariant. Furthermore, we connect the transformation of the source position directly with the local convergences around the image positions in Equation~\eqref{eq:gmst}.

\section{Conclusion}
\label{sec:conclusions}

\begin{figure*}[t]
\centering
  \includegraphics[width=0.63\textwidth]{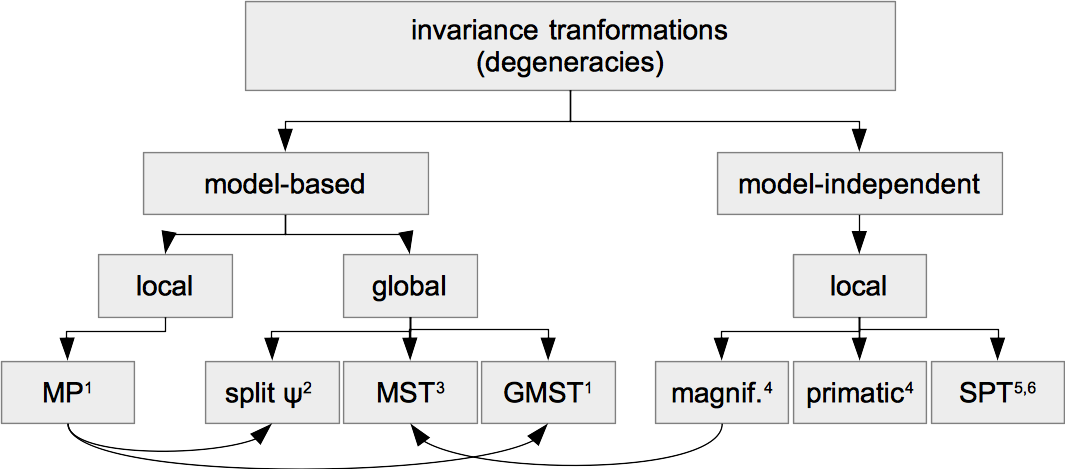}
    \caption{Summary of invariance transformations and degeneracies and their connections discussed here: the transformations in the local model-independent branch of the tree are set up as global transformations in the original works, yet, they have local restrictions that are intrinsic in the gravitational lensing formalism. $^1$ monopoles and generalised mass sheet transform by \cite{bib:Liesenborgs1}, $^2$ sub-division of the deflection potential discussed in \cite{bib:Wagner3}, $^3$ mass sheet transformation by \cite{bib:Falco} here listed under the model-based transformations as it is usually interpreted as a mass sheet but it could also be considered as a special case of the magnification transformation, $^4$ magnification transformation and prismatic transformation by \cite{bib:Gorenstein}, $^5$ source position transformation by \cite{bib:Schneider} with $^6$ time delay investigation by \cite{bib:Wertz}.}
\label{fig:summary}
\end{figure*}

\noindent
In this fourth part of the paper series of \cite{bib:Wagner1}, \cite{bib:Wagner2}, \cite{bib:Wagner3}, we derived that the standard single-lens-plane effective lensing formalism only allows for a very limited class of transformations, namely those constrained by Equation~\eqref{eq:invariance_transform}, to leave all observables invariant, given a cosmological model to fix the distances to the lens and the source. These most general invariance transformations are local transformations that leave the time delay differences, the relative image positions, the relative shapes, and the magnification ratios invariant. We found that the convergence cannot be altered at more than a null-set of positions, which can be explained by the fact that the solution of the Poisson equation requires a contiguous region to connect the two multiple images with the observed time delay difference. 

Without time delay information, Equation~\eqref{eq:gmst} defines the most general invariance transformation that leaves all remaining observables invariant. It is a local version of the SPT and connects the source position transformation with the local convergences at the positions of the multiple images. Contrary to the case in which time delay information is available, the transformations of the individual convergences in the regions around the multiple images are highly degenerate, as they only appear as part of the integrand over the regions around the multiple images.

We are interested in those local transformations confined to the areas of observed multiple images because, as discovered in \cite{bib:Wagner0} and \cite{bib:Tessore}, these are the only regions, in which we can constrain lens properties by observables without assuming a lens model. 
Taking a look at the model-independent constraints as derived in \cite{bib:Wagner1}, \cite{bib:Wagner2}, \cite{bib:Wagner3}, we find that the ratios of potential derivatives on the left-hand side are determined by observable, and thus invariant, quantities on the right-hand side. From this, we conclude that no further invariance transformations than those mentioned above exist that can be applied to these equations. 

We have briefly summarised the already existing classes of known invariance transformations as introduced in \cite{bib:Falco}, \cite{bib:Gorenstein}, \cite{bib:Saha}, \cite{bib:Liesenborgs1}, and \cite{bib:Schneider}, and integrated them as special cases into our derivations. All of them consider global invariance transformations and investigate their impact on lens models. As lens models make predictions about the mass density distribution or the deflection potential in regions without constraining multiple images, they are subject to additional, model-based degeneracies that arise in these regions and that do not occur in our purely data-based, model-independent approach. Figure~\ref{fig:summary} summarises the links between the different invariance transformations and degeneracies we discussed.

\begin{acknowledgements}
I would like to thank Matthias Bartelmann, Michael F. Herbst, Jori Liesenborgs, Sven Meyer, Dominique Sluse, Sebastian Stapelberg, R\"{u}diger Vaas, Olivier Wertz, and the anonymous referee for helpful comments, and Marc Gorenstein for his inspiring paper. I gratefully acknowledge the support by the Deutsche Forschungsgemeinschaft (DFG) WA3547/1-3.
\end{acknowledgements}

\bibliographystyle{aa}
\bibliography{aa}

\appendix

\section{Derivation of $\Delta \mathcal{G}(\boldsymbol{\alpha},\boldsymbol{x}_i, \boldsymbol{x_j})$}
\label{app:alpha}

We abbreviate $\boldsymbol{\alpha}(\boldsymbol{x}_i)$ by $\boldsymbol{\alpha}_i = (\alpha_{i1}, \alpha_{i2})$.
Inserting Equation~\eqref{eq:lens_mapping} into Equation~\eqref{eq:time_delay}, we derive
\begin{align}
\Delta \mathcal{G}(\boldsymbol{\alpha},\boldsymbol{x}_i, \boldsymbol{x_j}) &= \dfrac12 \left( \boldsymbol{\alpha}_i^2 - \boldsymbol{\alpha}_j^2 \right) \\ 
&= \dfrac12 \left( |\boldsymbol{\alpha}_i|^2 - |\boldsymbol{\alpha}_j|^2 \right) \\
&= \dfrac12 \left( (\alpha_{i1}^2 + \alpha_{i2}^2) - (\alpha_{j1}^2 + \alpha_{j2}^2) \right) \\
&= \dfrac12 \left( (\alpha_{i1} - \alpha_{j1})( \alpha_{i1} + \alpha_{j1})+ \right.\nonumber \\
& \qquad \left. (\alpha_{i2} - \alpha_{j2})(\alpha_{i2} + \alpha_{j2}) \right) \;.
\label{eq:component_wise}
\end{align}
Next, we note that
\begin{equation}
\boldsymbol{x}_i - \boldsymbol{\alpha}_i = \boldsymbol{x}_j - \boldsymbol{\alpha}_j \quad \Leftrightarrow \quad \boldsymbol{x}_i - \boldsymbol{x}_j = \boldsymbol{\alpha}_i - \boldsymbol{\alpha}_j \;,
\end{equation}
and insert the individual components into Equation~\eqref{eq:component_wise} to obtain
\begin{align}
\Delta \mathcal{G}(\boldsymbol{\alpha},\boldsymbol{x}_i, \boldsymbol{x_j}) &= \dfrac12 \left( (x_{i1} - x_{j1})( \alpha_{i1} + \alpha_{j1})+ \right. \nonumber \\
& \qquad \left. (x_{i2} - x_{j2})(\alpha_{i2} + \alpha_{j2}) \right) \\
&= \dfrac12 (\boldsymbol{x}_{i} - \boldsymbol{x}_{j})^\top ( \boldsymbol{\alpha}_{i} + \boldsymbol{\alpha}_{j})\;,
\end{align}
which is Equation~\eqref{eq:time_delay_alpha}.

\section{Proof for $\delta \kappa (\boldsymbol{x}) = 0$}
\label{app:ftc}

We begin with Equation~\eqref{eq:poisson_equation} for the original and the transformed deflection potential
\begin{align}
\text{L} \left(\psi(\boldsymbol{x})\right) &= 2 \kappa(\boldsymbol{x}) \;, \\
\text{L} \left( \tilde{\psi}(\boldsymbol{x})\right) &= 2 \tilde{\kappa}(\boldsymbol{x}) \;,
\end{align}
from which follows
\begin{align}
\text{L} \left( \delta \psi(\boldsymbol{x}) \right)&= 2 \delta \kappa(\boldsymbol{x}) 
\end{align}
by subtracting the two equations from each other and using the linearity of the Laplace operator.

Employing Green's theorem under the assumptions for the convergence stated in Section~\ref{sec:ti_kappa}, it follows that
\begin{align}
\delta \psi(\hat{\boldsymbol{x}}) &= \dfrac{1}{\pi} \int_{\mathcal{X}} \mathrm{d}^2 \boldsymbol{x}  \, \delta \kappa(\boldsymbol{x}) \ln \left| \hat{\boldsymbol{x}} - \boldsymbol{x} \right| \;,
\end{align}
which has to hold in $\overline{\mathcal{X}}$. Analogously,
\begin{align}
\delta \boldsymbol{\alpha}(\hat{\boldsymbol{x}}) &= \dfrac{1}{\pi} \int_{\mathcal{X}} \mathrm{d}^2 \boldsymbol{x}  \, \delta \kappa(\boldsymbol{x}) \dfrac{\hat{\boldsymbol{x}}-\boldsymbol{x}}{\left|\hat{\boldsymbol{x}}-\boldsymbol{x} \right|^2}
\end{align}
must hold for the deflection angle in $\overline{\mathcal{X}}$.
We insert the image positions $\boldsymbol{x}_i$ and $\boldsymbol{x}_j$ into these definitions and subsequently replace the respective expressions in Equation~\eqref{eq:invariance_transform} by the results
\begin{align}
\delta \psi(\boldsymbol{x}_i) - \delta \psi(\boldsymbol{x}_j) &= \dfrac{1}{\pi} \int_{\mathcal{X}} \mathrm{d}^2 \boldsymbol{x}  \, \delta \kappa(\boldsymbol{x}) \ln \left( \dfrac{\left| \boldsymbol{x}_i - \boldsymbol{x} \right|}{ \left| \boldsymbol{x}_j - \boldsymbol{x} \right|} \right) \;, \\
\delta \boldsymbol{\alpha}(\boldsymbol{x}_i) +  \delta \boldsymbol{\alpha}(\boldsymbol{x}_j) &=   \dfrac{1}{\pi} \int_{\mathcal{X}} \mathrm{d}^2 \boldsymbol{x}  \, \delta \kappa(\boldsymbol{x}) \left( \dfrac{\boldsymbol{x}_i-\boldsymbol{x}}{\left|\boldsymbol{x}_i-\boldsymbol{x} \right|^2} + \dfrac{\boldsymbol{x}_j-\boldsymbol{x}}{\left|\boldsymbol{x}_j-\boldsymbol{x} \right|^2} \right) \;.
\end{align}
Rearranging the terms, we arrive at Equation~\eqref{eq:invariance_transform_kappa2}. From it follows that
\begin{align}
\delta \kappa(\boldsymbol{x})  \, \mathrm{G}_1\left(\boldsymbol{x}_i, \boldsymbol{x}_j, \boldsymbol{x} \right) = \delta \kappa(\boldsymbol{x})  \, \mathrm{G}_2\left(\boldsymbol{x}_i, \boldsymbol{x}_j, \boldsymbol{x} \right) \;,
\end{align}
almost everywhere according to the definition of the Lebesgue integral. As a consequence, $\delta \kappa(\boldsymbol{x}) = 0$ must hold almost everywhere, given $G_1\left(\boldsymbol{x}_i, \boldsymbol{x}_j, \boldsymbol{x} \right)$ and $G_2\left(\boldsymbol{x}_i, \boldsymbol{x}_j, \boldsymbol{x} \right)$ as defined in Section~\ref{sec:ti_kappa}.

While we treated the special case of $g(\boldsymbol{x}) = 0$ on $\partial \mathcal{X}$ to set up Equations~\eqref{eq:potential} and \eqref{eq:angle}, Equation~\eqref{eq:invariance_transform} is also valid for any $g(\boldsymbol{x}) \in C^0(\overline{\mathcal{X}})$, because the boundary terms for $\psi(\hat{\boldsymbol{x}})$ amount to
\begin{align}
\int_{\partial \mathcal{X}} \mathrm{d}l \; \psi(\boldsymbol{x}) \partial_\nu \ln \left(|\hat{\boldsymbol{x}} - \boldsymbol{x}|\right) - \int_{\partial \mathcal{X}} \mathrm{d}l \; \ln \left(|\hat{\boldsymbol{x}} - \boldsymbol{x}|\right) \partial_\nu \psi(\boldsymbol{x}) \;,
\end{align}
in which $\mathrm{d}l$ denotes the infinitesimal boundary element and $\partial_\nu$ the outer normal derivative. Hence, calculating $\delta \psi (\hat{\boldsymbol{x}})$ and $\delta \boldsymbol{\alpha}(\hat{\boldsymbol{x}})$, they vanish under Dirichlet boundary conditions and the assumption that the values of $\psi(\boldsymbol{x})$ given at the boundary are observables and thus invariant under the transformation.

\end{document}